 \documentclass[final,5p,times,twocolumn]{elsarticle}

\usepackage{amssymb}

\journal{Icarus}

\bibpunct{[}{]}{;}{a}{,}{,}
\usepackage{graphicx}
\usepackage{natbib}
\newcommand{\degr}{\ensuremath{^\circ}}
\newcommand{\arcsec}{\mbox{\ensuremath{^{\prime\prime}}}}

\usepackage{hyperref}
\hypersetup{
   bookmarks=true,         
   unicode=false,          
   pdftoolbar=true,        
   pdfmenubar=true,        
   pdffitwindow=false,     
   pdftitle={First disk-resolved spectroscopy of (4) Vesta},
   pdfauthor={Benoit Carry, P. Vernazza, C. Dumas and M. Fulchignoni}, 
   pdfsubject={Planetary Science}, 
   pdfkeywords={},         
   pdfnewwindow=true,      
   colorlinks=true,        
   linkcolor=gray,         
   citecolor=blue,         
   filecolor=gray,         
   urlcolor=gray           
}

 \graphicspath{{figures/}}

\newcommand{\gJ}{\textsl{J}}
\newcommand{\gHK}{\textsl{H}+\textsl{K}}

\usepackage{pstricks}

\begin{document}

\begin{frontmatter}



\title{First disk-resolved spectroscopy of (4) Vesta\tnoteref{1}}
\tnotetext[1]{Based on observations
 collected at the European Southern Observatory, Paranal, Chile -
 \href{http://archive.eso.org/wdb/wdb/eso/sched_rep_arc/query?progid=60.A-9041(A)}%
      {60.A-9041}}


\author[lesia,eso]{Beno\^{i}t Carry}
\author[esa]{Pierre Vernazza}
\author[eso]{Christophe Dumas}
\author[lesia]{Marcello Fulchignoni}

\address[lesia]{LESIA, Observatoire de Paris-Meudon, 5 place Jules
 Janssens, 92195 Meudon Cedex, France} 
\address[eso]{ESO, Alonso de C\'{o}rdova 3107, Vitacura, Casilla
 19001, Santiago de Chile, Chile} 
\address[esa]{Research and Scientific Support Department, European
 Space Agency, Keplerlaan 1, 2201 AZ Noordwijk, The Netherlands} 

\begin{abstract}
 \indent Vesta, the second largest Main Belt asteroid, will be the first
 to be explored in 2011 by NASA's Dawn mission.
 It is a dry, likely differentiated body with spectrum suggesting that is has been
 resurfaced by basaltic lava flows, not too different from the lunar maria. \\
 \indent Here we present the first disk-resolved spectroscopic
 observations of an asteroid from the ground. We observed (4) Vesta 
with the ESO-VLT adaptive optics equipped
 integral-field near-infrared spectrograph 
 SINFONI, as part of its science verification campaign.
 The highest spatial resolution of $\sim$90 km on Vesta's surface was
 obtained during excellent seeing conditions (0.5\arcsec) in October
 2004. \\ 
 \indent We observe spectral variations 
 across Vesta' surface that can be interpreted as variations of either 
 the pyroxene composition, or the effect of surface aging.
 We compare Vesta's 2 micron absorption band to that of
 howardite-eucrite-diogenite (HED) meteorites
 that are thought to originate from
 Vesta, and establish particular links between specific regions
 and HED subclasses.
 The overall composition is found to be mostly compatible with
 howardite meteorites, although a small area around 180\degr~East longitude
 could be attributed to a diogenite-rich spot.
 We finally focus our spectral analysis on the characteristics of Vesta's bright and dark regions
 as seen from Hubble Space Telescope's visible and
 Keck-II's near-infrared images. 

\end{abstract}


\end{frontmatter}



\section{Introduction}
 \indent Vesta, with a mean radius of 265 $\pm$ 5 km
 \citep{1997-Science-277-Thomas, 1997-Icarus-128-Thomas}, 
 is the second largest Main Belt asteroid. It orbits the
 Sun at a semi-major axis of $a = 2.36$ AU. It is the only
 known differentiated asteroid with an intact internal structure,
 presumably consisting of a metal core, an ultramafic mantle, and a
 basaltic crust \citep[see ][for a review]{2002-AsteroidsIII-4.3-Keil}.
 The igneous nature of its surface material was diagnosed in the early seventies
 \citep{1970-Science-168-McCord} and subsequently confirmed by
 additional observations \citep{1975-Icarus-26-Larson,1977-Icarus-31-McFadden,1997-Icarus-128-Binzel, 
   1997-Icarus-127-Gaffey,2005-AA-436-Vernazza}.
 Two decades ago, Vesta was still the only known
 asteroid with a basaltic surface
 until \citet{1993-Science-260-Binzel} discovered several main-belt
 asteroids with diameters $<$10 km and surface composition
 similar to Vesta's \citep[see][for a review]{2006-ACM-Pieters}.
 Those V-type asteroids were identified as members of Vesta's dynamical
 family \citep[the so-called Vestoids:][]{1989-AsteroidsII-6.3-Williams,1990-AJ-100-Zappala, 
 1996-AA-316-Marzari}. The hypothesis of a large collision on Vesta
 \citep{1996-AA-316-Marzari} has been confirmed by Hubble Space
 Telescope (HST) observations revealing the presence of a $\sim$460 km
 basin on its surface  
 \citep[1.7$\times$ Vesta's radius,][]{1997-Science-277-Thomas},
 which is most likely the result of an impact 
 with a $\sim$35 km projectile \citep{1997-MPS-32-Asphaug}.
 Several other V-type asteroids have seen been discovered among the
 Near-Earth Asteroids (NEA) population \citep{1995-Icarus-115-Xu,
 2004-Icarus-170-Binzel,2005-Icarus-175-Marchi-II}, and main-belt
 (1459 Magnya with a semi-major axis of 3.14 AU
 \citep{2000-Science-288-Lazzaro} and some others in the middle belt
 \citep{2008-Icarus-198-Moskovitz}).  \\
 \indent Spectroscopic observations at visible \citep{1970-Science-168-McCord}
 and near-infrared (NIR) \citep{1975-Icarus-26-Larson, 1980-GeCoA-44-Feierberg, 1980-Science-209-Feierberg}
 wavelengths revealed that the disk-integrated spectrum of Vesta displays strong similarities with
 laboratory spectra of howardite, eucrite and diogenite (HED) meteorites,
\citet{1970-Science-168-McCord} implying quite naturally that HED meteorites came
 from Vesta. Later, Earth-based \citep{1997-Icarus-127-Gaffey,2005-AA-436-Vernazza} 
 rotationally resolved disk-integrated spectrophotometric measurements 
 confirmed a geologically heterogeneous surface \citep[\textsl{e.g.}][]{1983-LPI-14-Gaffey},
 consistent with the composition of HED meteorites
 \citep[see the review by][]{2002-AsteroidsIII-4.3-Keil}.

 \indent Multiband disk-resolved imaging at four different visible
wavelengths (0.439, 0.673, 0.953 and 1.042 $\mu$m) with HST
 \citep{1997-Icarus-128-Zellner} have revealed various high contrast 
 albedo marks across its surface \citep{1997-Icarus-128-Binzel}, also
 observed by Keck in the near-IR \citep[2 and 3.6 $\mu$m,][]{2005-Icarus-177-Zellner}. \\
 \indent Several scenarios have been proposed to explain these strong
 albedo variations. \citet{1997-Icarus-128-Binzel} suggested differences in mineralogy, 
 grain size, or space weathering processes. \citet{2005-Icarus-177-Zellner} discussed that 
 Vesta's dark regions may be ``impact basins'' later filled with basaltic lavas, much like 
the lunar maria. Interestingly, \citet{1997-Science-277-Thomas} found a correlation between 
the width and depth of the 1 micron spectral band and the togography
(with respect to Vesta's mean 
 elevation) in regions near the South pole crater, supporting
 differentiation of Vesta mineralogy  
 throughout its crust. \\ 
 \indent While much attention has been given to the mineralogical
 characterization of Vesta' surface, little has been said about its
 color. The surface properties of Vesta -- mainly the characteristics
 of its spectral signatures and relatively high albedo --  
 suggest that its surface is somewhat protected from heavy space
 weathering \citep{1995-Icarus-115-Hiroi}. 
However, recent laboratory experiments demonstrate that stronger
 spectral differences between Vesta and the HED meteorites
 should be expected \citep{2005-AA-443-Marchi-I,2006-AA-451-Vernazza}.
\citet{2005-AA-443-Marchi-I} and  \citet{2006-AA-451-Vernazza}
performed ion irradiation 
 experiments on pyroxene and eucrite samples and found that their reflectance
 spectrum reddens with progressive irradiation. 
 As those minerals are the main components of Vesta' surface, one 
 would expect solar-wind ions to redden its spectrum and lower its albedo.  \\
 \indent The spectroscopic match between Vesta and the HED meteorites
 implies that either some processes rejuvenate its surface continuously, or
 that the action of the solar wind ions onto its surface is lower than expected. 
 From the dynamical dispersion of the Vestoid members, we can infer that the impact 
 event responsible of the southern crater has occurred more than 1 Gyr ago
 \citep{1996-AA-316-Marzari, 2008-Icarus-193-Nesvorny}, which is longer than the timescale
 needed for space weathering to alter Vesta' surface properties
 \citep{2006-Icarus-184-Brunetto, 2006-AA-451-Vernazza, 2009-Nature-458-Vernazza}. 
 Recently, \citet{2008-LPI-39-Shestopalov} proposed that continuous
 bombardment  of Vesta's surface by small meteors could be a possible mechanism 
 to rejuvenate its surface. They propose that numerous small-sized debris
 ejected during the large southern impact may have remained within the gravitational 
 sphere of influence of Vesta. Close encounters with other minor planets would then trigger 
 instabilities resulting in those fragments to fall onto Vesta, thus
 mixing its regolith and 
 erasing the space weathering effects. However, the quantity of debris and their orbital 
 timescale remain to be evaluated.
 \citet{2006-AA-451-Vernazza} suggested an alternative mechanism. 
 They showed that a remnant magnetic field of only $\sim$0.2
 micro-Tesla ($\mu$T) could shield a large portion of Vesta' surface from solar
 wind ions, explaining at once Vesta's largely unweathered 
 aspect and strong color variations (from the presence of magnetic ``cusps'').\\
 \indent Here we report the first ground-based disk-resolved spectral study
 of an asteroid surface. The near-IR mapping of Vesta' spectral slope and absorption bands
 properties allowed us to investigate possible causes for their spatial variations, such as 
 heterogeneity in its surface composition or in the strength of space
 weathering
 \citep{2006-Icarus-184-Brunetto}.


\section{Observations and Data Reduction}
 \subsection{SINFONI}

\begin{figure}
\begin{center}

 \textbf{Spectro-cube of (4) Vesta}\\

 \includegraphics[width=.5\textwidth]{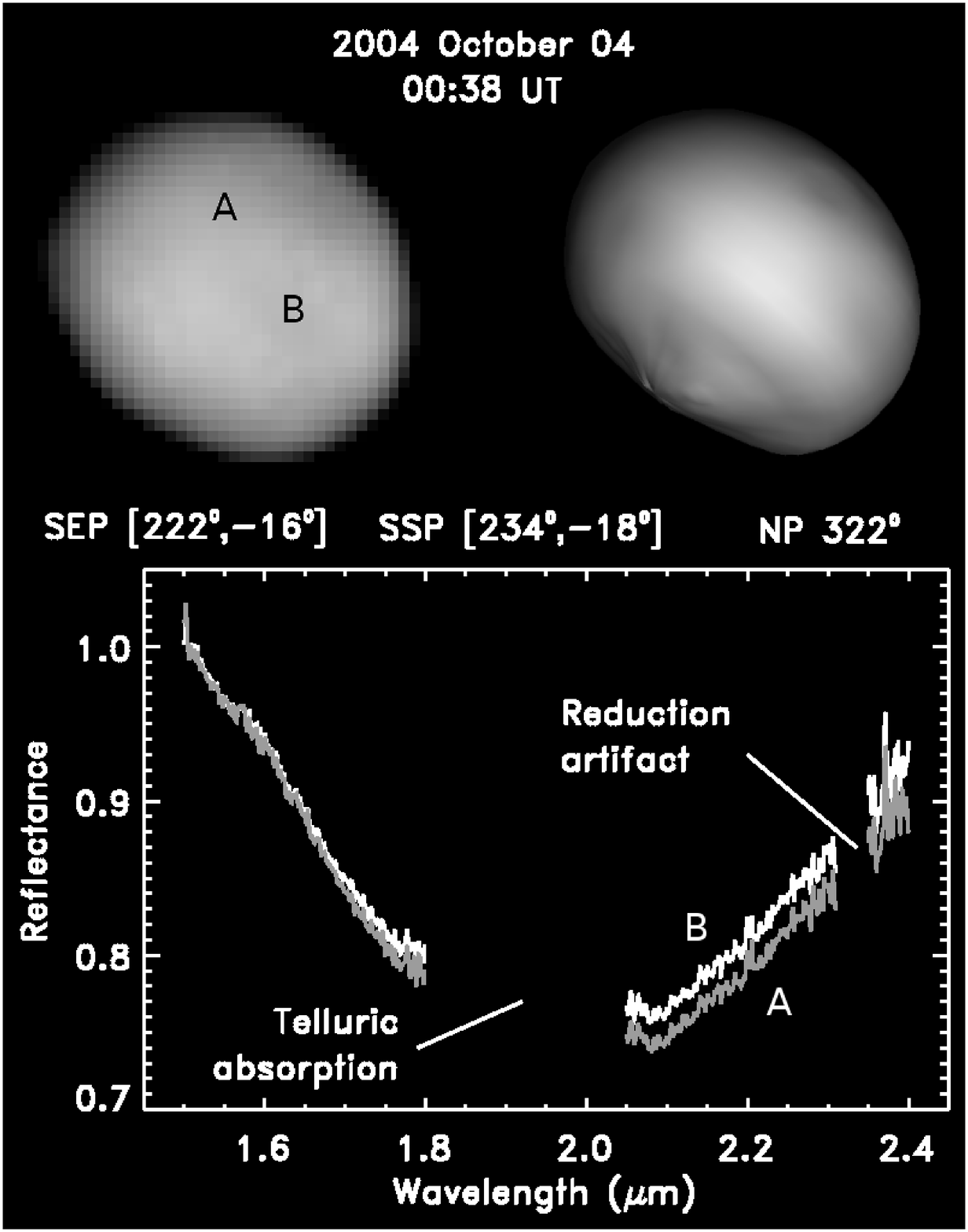}

 \caption[Spectro-cube of Vesta]{%
   \textsl{Top left:} Image of Vesta
   obtained with the \gHK~grating for the 2004 October 04 UT run, by 
   stacking all images of the cube along the wavelength direction. 
   Orientation is usual with North up and East left.
   \textsl{Top right:} The shape model of Vesta at the same orientation 
   \citep[as defined by][]{1997-Icarus-128-Thomas} is given 
   for comparison (model obtained from the Eproc ephemeris 
   generator (\texttt{http://www.imcce.fr})). 
   We report the Sub-Earth Point (SEP) and Sub-Solar Point (SSP)
   coordinates (longitude, latitude) as well as the pole angle (NP:
   defined as the angle in the plane of the sky between the celestial
   north and the projected asteroid spin-vector, measured
   counter-clockwise, from North through East).\\
   \textsl{Lower panel:} \gHK~spectra of two arbitrary selected
   spaxels A and B. We normalized all the spectra to unity at 1.5 $\mu$m within
   this range and we degraded the spectral resolution to improve the
   signal-to-noise ratio by smoothing the spectra with a 8-pixel box median
   filter. We removed some artefacts around 2.33 $\mu$m that contamnated 
   our image-cubes.\\
   This figure illustrates the peculiarity of a spectro-cube: 
   it is a three dimensional array where each
   spaxel in the spatial plane is a spectrum in the third dimension.
   \label{fig-cube}
   \label{lastfig}
 }
 \end{center}
\end{figure}

   \indent Vesta was observed with the Spectrograph for INtegral Field Observations in the Near
   Infrared (SINFONI) \citep{2004-Msngr-117-Bonnet} during the 2004 science verification campaign 
   (Prog. ID: 60.A-9041) of the instrument. SINFONI is mounted on the cassegrain focus of the 
   Yepun telescope (UT4) at the European Southern Observatory's (ESO)
   Very Large Telescope (VLT).
   SINFONI's main sub-systems are the Adaptive Optics
   (AO) module (Multi-Application Curvature Adaptive Optics: MACAO)
   developed at ESO \citep{2003-SPIE-4839-Bonnet}, 
   and the integral-field unit (SPectrometer for Infrared Faint Field Imaging: SPIFFI)
   developed at the Max Planck Institute for Extraterrestrische Physik (MPE)
   in collaboration with Nederlandse Onderzoekschool Voor Astronomie (NOVA) and ESO
   \citep{2003-SPIE-1548-Eisenhauer}. An ``image slicer'' cuts the field of view (FoV) into 32
   image-slitlets, which are redirected towards the spectrograph grating, to be re-imaged onto 
   a $2048 \times 2048$ pixels Hawaii 2RG detector. 
   The original FoV is reconstructed into an 64x64 pixel image-cube, each 
   slice of the cube corresponding to an image of the FoV at a given wavelength.   
   Fig.~\ref{fig-cube} presents an example of a reconstructed image.
%
%
%
%
%
 \subsection{Observations}
   We observed Vesta at three epochs: 2004 August 21,
   August 23 and October 4 (Table~\ref{tab-obs-condition}), 
   one month apart from Vesta's opposition (2004, September 13 UT).
   Data obtained on August 23 were impacted by poor seeing
   conditions and were thus discarded from our analysis.   
   We used the highest angular-resolution provided by SPIFFI with an 
   equivalent pixel size on sky of $25\times12.5$ milli-arcsec
   and a resulting field of view of $0.8 \times 0.8$ square arcseconds.
   Image-cubes were obtained across the [1.1-2.4] $\mu$m range with the
   \gJ~and \gHK~gratings (Table~\ref{tab-obs-settings}).
%
%
%
%
%
%
\begin{table*}
\begin{center}

 \textbf{Observing Conditions}\\
 \begin{tabular}{cccccccccccc}
  \hline\hline
   Date & Grating & $\Delta$ & $r$ & $\alpha$ & $\phi$ & SEP$_\lambda$ & SEP$_\varphi$ & X & Seeing &\multicolumn{2}{c}{$\Theta^{\dagger}$}\\
     (UT)    &        & (AU) & (AU) & (\degr) & (\arcsec) & (\degr) & (\degr)  & &(\arcsec)  & (\arcsec) & (km) \\ 
   \hline
   2004 Aug 21 - 04:59 & \gHK & 2.35 & 1.41 & 12 & 0.55 & 179 & -15 & 1.13 & 1.1 & 0.084 & 144 \\
   2004 Aug 21 - 08:28 & \gJ  & 2.35 & 1.41 & 12 & 0.55 & 304 & -15 & 1.14 & 1.1 & 0.218 & 371 \\
   2004 Aug 21 - 08:46 & \gHK & 2.35 & 1.41 & 12 & 0.55 & 285 & -15 & 1.18 & 1.2 & 0.084 & 144 \\
   2004 Oct 04 - 00:20 & \gJ  & 2.39 & 1.45 & 11 & 0.53 & 240 & -15 & 1.34 & 0.5 & 0.053 & 92 \\ 
   2004 Oct 04 - 00:33 & \gHK & 2.39 & 1.45 & 11 & 0.53 & 226 & -15 & 1.29 & 0.5 & 0.055 & 95 \\ 
  \hline
 \end{tabular}

 \caption[Observing Conditions]{%
   Vesta's heliocentric ($\Delta$) and geocentric ($r$) distances,
   phase angle ($\alpha$),
   angular diameter ($\phi$) and
   Sub-Earth Point (SEP) coordinates$^{\star}$
   (longitude $\lambda$ and latitude $\varphi$)
   for each epoch (given in UT).
   Airmass (X), seeing and corresponding 
   resolution element ($\Theta$) at 1.2 and 2.2 micron
   (for the \gJ~and \gHK~grating respectively)
   in arcseconds on sky and
   kilometers on Vesta's surface,
   are also reported for each observation.
   The poor spatial resolution obtained with the \gJ~grating
   in August is due to the poor meteo conditions that prevailed at that time.
   The data are available since December 2004 on the
   \href{http://archive.eso.org/eso/eso_archive_main.html}%
        {ESO archive}, under the program ID: 60.A-9041.\\
   $^{\dagger}$evaluated on the solar analog spectro-cubes.\\
   $^{\star}$coordinate system is planetocentric \citep[IAU
     recommendation:][]{2007-CeMDA-98-Seidelmann}

   \label{tab-obs-condition}
   \label{lasttable}
 }
\end{center}
\end{table*}
%
%
%
%
%
%
\begin{table}
\begin{center}

 \textbf{Instrumental Settings}\\
 \begin{tabular}{ccccc}
  \hline\hline
  Grating & $\lambda$  & $\lambda$/$\Delta \lambda$ & Pixel scale & FoV    \\
          &  ($\mu$m)  &                            &  (mas/pix)   & (arcsec$^{2}$)\\
  \hline
  \gJ  & 1.10-1.40 & $\sim$3000 & 25$\times$12.5 & 0.8$\times$0.8 \\
  \gHK & 1.45-2.45 & $\sim$1500 & 25$\times$12.5 & 0.8$\times$0.8 \\
  \hline
 \end{tabular}

 \caption[Instrumental Settings]{
   Wavelength range ($\lambda$), spectral resolution (
   $\lambda$/$\Delta \lambda$), pixel scale (in milli-arcsecond per
   pixel) and size of the field of view (FoV) for each
   spectroscopic mode used in this study.

   \label{tab-obs-settings}
   \label{lasttable}}

\end{center}
\end{table}
%
%
%
%
\begin{table*}
\begin{center}

 \textbf{Solar Analogs}\\
 \begin{tabular}{cccccccccc}
  \hline\hline
   Designation & RA & DEC & Spectral& Mv & Date & Filter & Airmass  \\
    & (hh$:$mm$:$ss) & (dd$:$mm$:$ss) &Type &(mag.)&(UT) &    &    \\
  \hline
  HD 1835 & 00:22:52 & -12:12:34 & G5V & 6.4 & 2004 Aug 21 - 05:35 & \gHK & 1.10 \\
  HD 1835 & 00:22:52 & -12:12:34 & G5V & 6.4 & 2004 Aug 21 - 09:06 & \gJ  & 1.17 \\
  HD 1835 & 00:22:52 & -12:12:34 & G5V & 6.4 & 2004 Aug 21 - 09:13 & \gHK & 1.19 \\
  HD 1461 & 00:18:42 & -08:03:11 & G0V & 6.5 & 2004 Oct 04 - 00:54 & \gHK & 1.52 \\
  HD 1461 & 00:18:42 & -08:03:11 & G0V & 6.5 & 2004 Oct 04 - 01:04 & \gJ  & 1.45 \\
  \hline
 \end{tabular}

 \caption[Solar analogs]{%
   Solar analogs observed in this study, along with their equatorial coordinates, 
   spectral type and visual magnitude (Mv). The UT time of
   observation, the filter and the airmass are also reported. 
   \label{tab-obs-psf}
   \label{lasttable}
 }
\end{center}
\end{table*}
%
%
%
%
%
 \subsection{Data reduction\label{sec-reduction}}
   \indent We reduced the SINFONI data with the ESO data-reduction
   pipeline (version 1.8.2). At the time of the observations,
   the instrument was equipped with an
   engineering-grade detector \citep{2004-Msngr-117-Bonnet}, suitable
   for technical qualification but that
   had a large deffect affecting the data quality after 1.35
   and 2.4 $\mu$m for the \gJ~and
   \gHK~grating respectively. The \gJ~observations also presented higher noise
   at short wavelengths and we had to discard the 1.1-1.17 micron range.
   Distance and distortion calibration tables were provided by the
   commissioning team [H. Bonnet, private communication].
   Basic calibrations (dark frames, lamp flats, and Ar and Xe lamps for the
   wavelength calibration) were obtained via SINFONI
   calibration plan. Each objet-sky pair of frames were reduced using the pipeline 
   to correct from bad pixels, flat-fielding, sky subtraction and
   reconstruct the image-cubes. \\ 
   \indent Due to atmospheric differential refraction,
   the position of the object across the cube slowly drifts with the wavelength. 
   We thus re-aligned each slice of the image-cubes for both the solar analogs and Vesta.
   Solar analogs were also observed to correct our reflectance Vesta
   spectra from atmospheric absorption and solar color. 
   The spectra of Vesta, and of the solar analog observed at similar airmass (see
   Table~\ref{tab-obs-psf}), were then extracted using an aperture compromising
   the maximum flux collected and the minimum residual sky contribution. 
   Also, because the small FoV of SINFONI
   (0.8\arcsec$\times$0.8\arcsec) caused some light to be lost 
   in the wings of the PSF under poor seeing conditions, which had the effect of introducing
   an artificial slope in our solar analog spectra (see Fig.~\ref{fig:
     starslope}), we adjusted both the aperture size and 
   the overall spectral slope of our solar analog to match Vesta's
   disk-integrated spectrum \citep[from][]{2005-AA-436-Vernazza} 
   (Fig.~\ref{fig: vestaslope}).
   This process also permitted to correct for any
   artificial slope possibly introduced by different atmospheric
   extinction between 
   Vesta and the solar analogs observation.
   However, given this slope correction, we do not
   present absolute measurements of the spectral slope
   nor of Vesta's composition.
   But since we processed
   the different data sets in a similar way, they
   are \textsl{self-}consistent and we can use them to study
   variations on Vesta's surface.\\

%
%
\begin{figure}
\begin{center}
 \textbf{Slope of the Reference star}\\
 \includegraphics[width=.5\textwidth]{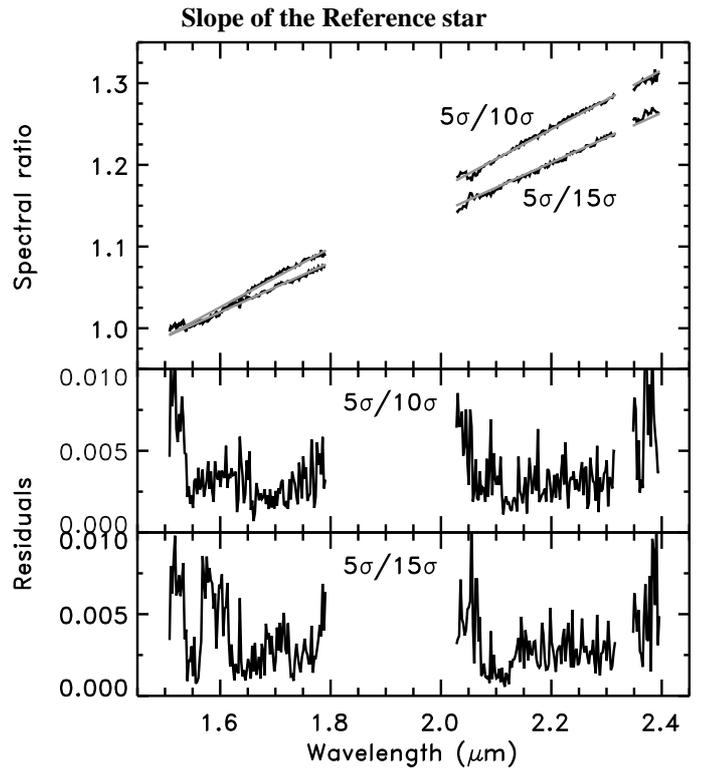}
 \caption[Slope of the Reference star]{
   Ratio of the reference star spectrum
   (HD 1835 obtained in August, see
   Table~\ref{tab-obs-psf}) to itself, for several apertures (corresponding to 5, 10 
   and 15 times the standard deviation of the star) over the \gHK~wavelength range.
   For each ratio, we computed its linear regression, and show the
   deviation (residuals) to this linear fit in the two lower panels.
   According to the chosen aperture a strong linear slope over the wavelength range is
   introduced. We corrected this effect by dividing the stellar spectra by linear
   function (see text), whose slope were adjusted to match previous observations 
   of Vesta \citep{2005-AA-436-Vernazza}.
 }
 \label{fig: starslope}
\end{center}
\end{figure}
\begin{figure}
\begin{center}
 \textbf{Vesta's slope for non-corrected stars}\\
 \includegraphics[width=.5\textwidth]{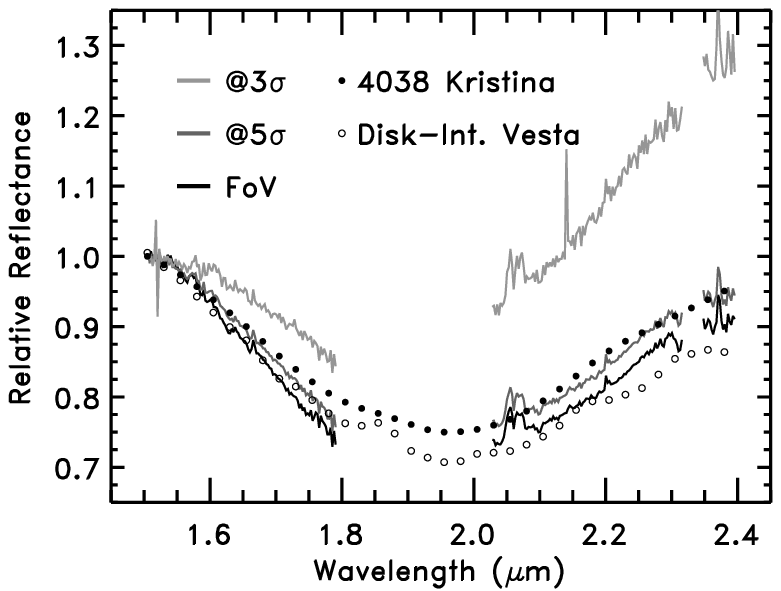}
 \caption[Vesta's slope for non-corrected stars]{
   Following Fig.~\ref{fig: starslope}, we show some disk-integrated spectra
   of Vesta obtained without the linear correction of the stellar
   spectra and for several aperture values (FoV meaning integrating the flux over the whole FoV).   
   We show for comparison in open circles the spectrum of Vesta
   obtained with long slit spectroscopy by \citet{2005-AA-436-Vernazza}, and in 
   filled circles the very red small Vestoid (4038) Kristina (see section~\ref{sec-SW} 
   for more info on this asteroid).
   This figure clearly shows the need for correcting  the stellar spectrum
   slope (see text).
 }
 \label{fig: vestaslope}
\end{center}
\end{figure}
\begin{figure}
\begin{center}
 \textbf{\gJ-band spectra}\\
 \includegraphics[width=.5\textwidth]{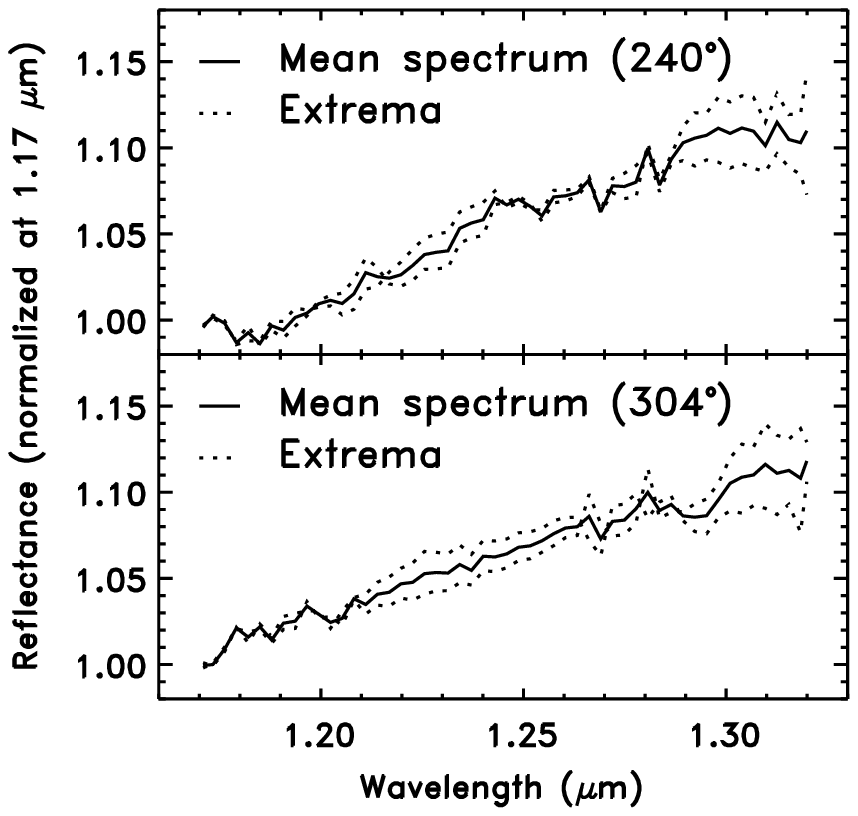}
 \caption[\gJ-band spectra]{%
   Average (solid line) and envelope spectra (dotted lines)
   obtained with the \gJ~grating
   normalized to unity at 1.17 $\mu$m.
   Each panel corresponds to a different rotational phase as
   reported on the figure
   (see Table~\ref{tab-obs-condition}).
   The slope heterogeneity across Vesta's surface is visible from the breadth
   of the envelops at long wavelengths 
     (see section~\ref{sec-slope}).
   \label{fig-specJ}
   \label{lastfig}
 }
\end{center}
\end{figure}
\begin{figure}
\begin{center}
 \textbf{\gHK~grating spectra}\\
 \includegraphics[width=.5\textwidth]{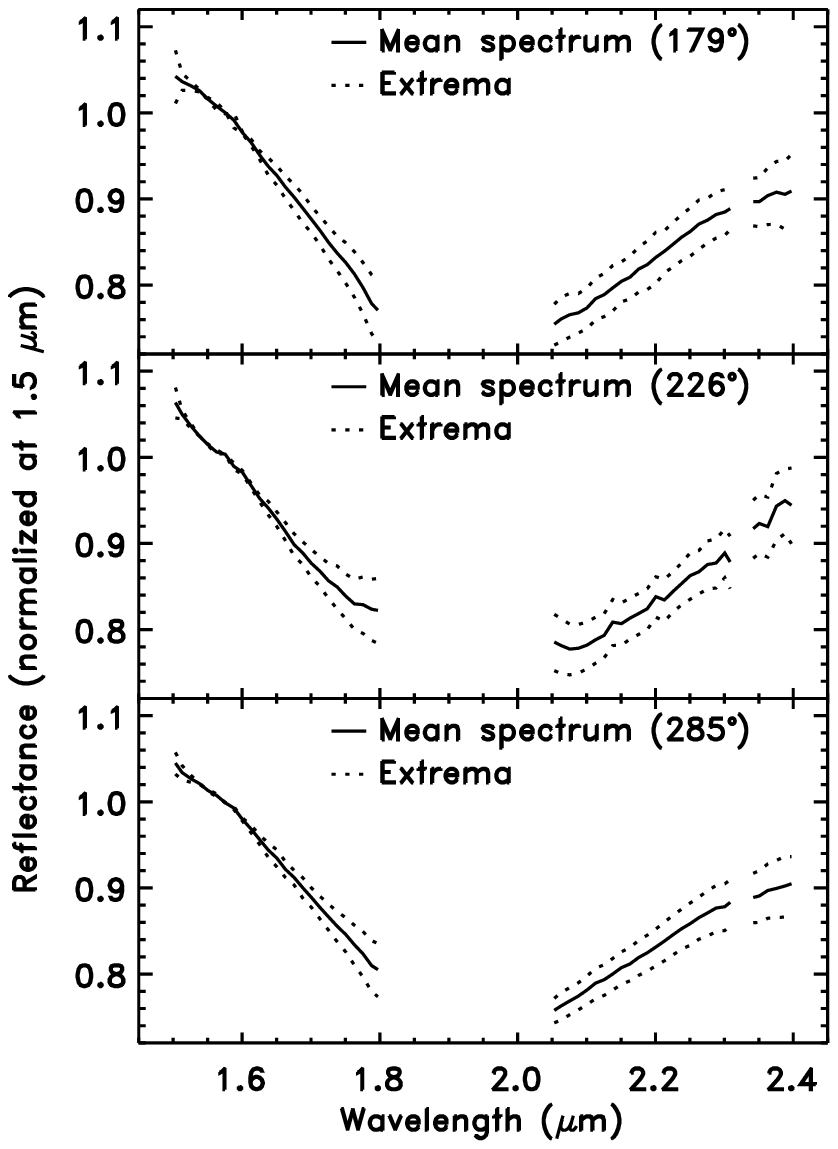}
 \caption[\gHK~grating spectra]{
   Average and envelope spectra
   obtained with the \gHK~grating
   normalized to unity at 1.5 $\mu$m.
   Each panel corresponds to a different rotational phase of
   Vesta as
   reported on the figure
   (see Table~\ref{tab-obs-condition}).
   Vesta's surface properties heterogeneity is visible from the breadth of the envelopes longward of 1.8 micron. .
   \label{fig-specHK}
   \label{lastfig}
 }
\end{center}
\end{figure}

%
%
%
%
%
   \indent As already addressed by several authors
   \citep[\textsl{e.g.}][]{1997-Icarus-128-Binzel},
   the grazing viewing angle affecting the region near Vesta's limb
   causes unreliable flux measurement for the outer annulus of the
   asteroid's apparent disk.
   We thus limited
   our study to a restricted region of interest (ROI)
   covering the innermost portion of the disk.
   The ROI was defined by Vesta's projected shape
   reduced to a given fraction \citep{2008-AA-478-Carry},
   corresponding to a ROI of 60\%
   of Vesta's radius.
   We show the spectra in both wavelength ranges
   for each rotational phase in Fig.~\ref{fig-specJ} and
   Fig.~\ref{fig-specHK}.
   Spectra are normalized to 
   unity at 1.17 $\mu$m and 1.5 $\mu$m for the \gJ~and
   \gHK~gratings respectively; 
   and smoothed with a median filter, using a box size of 8
   pixels in the spectral direction for each spaxel.\\
%
%
%
%
%
%


\section{Pyroxene Distribution\label{sec-pyroxene}}

\begin{figure}
\begin{center}
 \textbf{HK-band slope removed}\\
 \includegraphics[width=.5\textwidth]{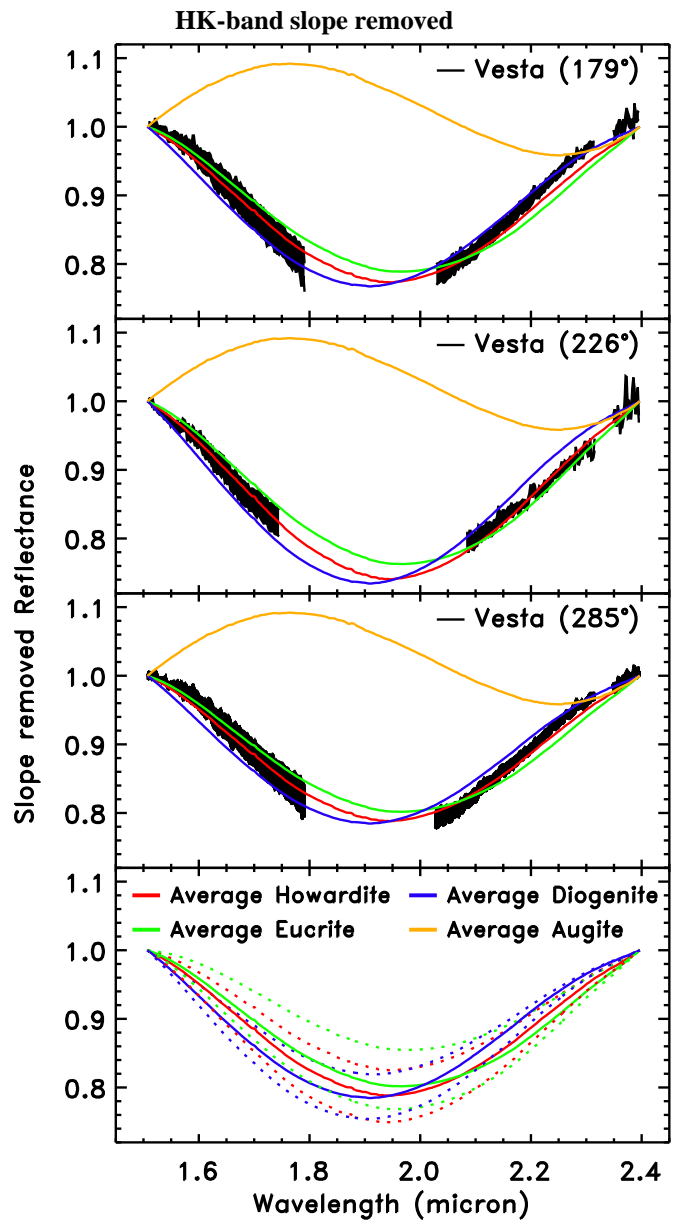}
 \caption[HK-band slope removed]{%
   Vesta's spectra for the three \gHK~observations after HK-band spectral slope
   removal (see text).
   For each data-set, we overplot the average HED and augite
   spectra (based on 20, 76, 20 and 6 samples respectively) after
   slope removal.
   The lowermost panel represents the HED meteorites spectra with their
   standard deviation shown as dotted lines.
   The latter one is strongly affected by
   slope and grain size hiding the influence of mineralogy.
   Although the classes overlap close to the band center, they
   are separable near the band ends.
   From the global band shape at each rotational phase, one can
   already see that the Vesta's spectra above will neither match pure
   eucrite nor pure diogenite composition, but a mix of these minerals instead. 
   \label{fig-bandHK}
   \label{lastfig}
 }
\end{center}
\end{figure}

\begin{figure}
\begin{center}
 \textbf{Band wings fitting method}\\
 \includegraphics[width=.5\textwidth]{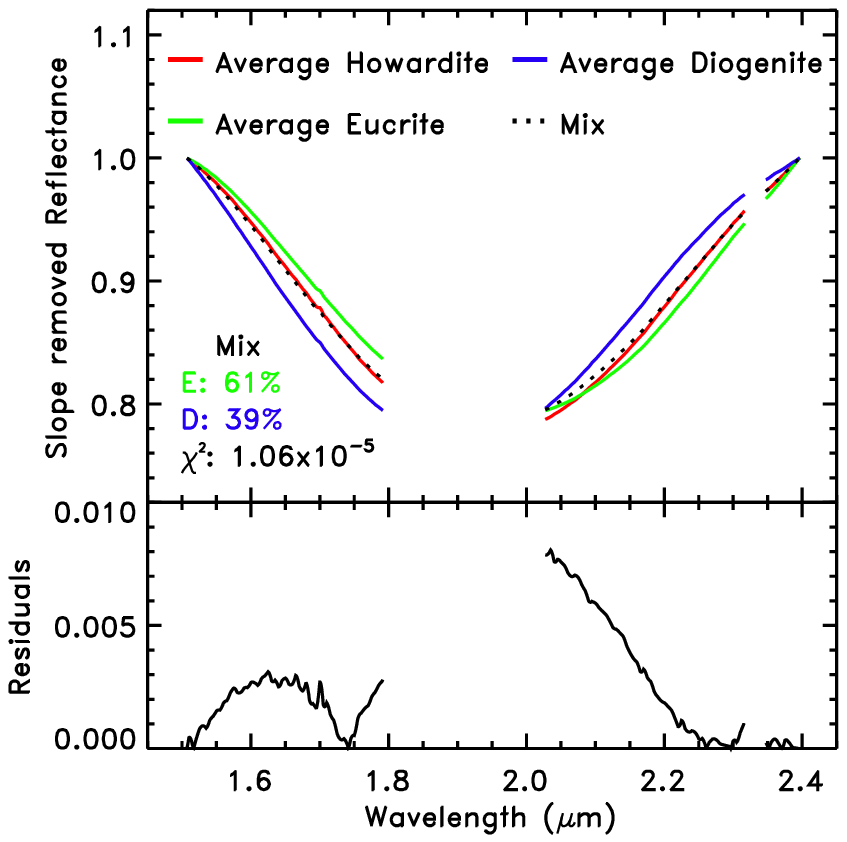}
 \caption[Band wings fitting method]{%
   We present here the method used to determine Vesta's mineralogy in
   terms of mixing rations of eucrites and diogenites. 
   The upper panel shows the average spectra of the howardite,
   eucrite and diogenite meteorites used in this study. We fit the
   howardite spectrum with a combination of 61\% of eucrite and 39\%
   of diogenite, The fit residuals are showed in the lower panel.
   Although the choice of end-members in our fit is likely not unique, the low
   level of residuals obtained indicates that this technique can be used as a good approximation to 
   study Vesta mineralogy. 
   \label{fig: howardite}
   \label{last-fig}
 }
\end{center}
\end{figure}

  \indent We present a mineralogical study of Vesta based on the analysis of 
  spatially resolved reflectance spectra in the \gHK~range, and by looking for 
  correspondence between regions of Vesta surface and  HED
  subclasses and  pyroxene minerals (orthopyroxene, clinopyroxene).\\
  \indent Pyroxene spectra are characterized by broad
  absorption bands centered at 1 and 2 $\mu$m. The wavelength 
  position of these absorption features constrains
  the nature of the minerals \citep{1974-JGR-79-Adams}. In our case, the
  telluric absorptions strongly affected the central region of the
  \gHK~range (Fig.~\ref{fig-cube}), preventing a detailed
  investigation of pyroxene compositional variations across 
  Vesta's surface. However, one can still perform first level characterization
  of the pyroxene composition as the intensity of the slopes of the upward and downward 
  2 $\mu$m absorption band are directly related to the detailed chemical content of the pyroxenes. 
  A gradual downward slope associated with a steep upward slope indicate a position of the 
  band center shifted towards short wavelengths. The absorption shifts towards short wavelengths
  for low calcium pyroxenes (\textsl{i.e.} orthopyroxene), while the opposite is true for 
  high calcium pyroxenes (\textsl{i.e.} clinopyroxene). For HED meteorites, the absorption 
  is positioned at short wavelengths for diogenites, and increases for the howardites 
  and eucrites \citep{1997-Icarus-127-Gaffey}. This trend appears clearly from
  average HED spectra: we computed the average
  spectra for the howardite, eucrite and diogenite meteorite
  classes, using 20, 20 and 76 samples (from
  RELAB: \href{http://www.planetary.brown.edu/relab/}%
  {\texttt{http://www.planetary.brown.edu/relab/}})
  respectively; we also computed 
  the average spectrum for augite (high calcium clinopyroxene, 6 samples). These
  mean spectra, scaled to unity at 1.5 $\mu$m, are shown in Fig.~\ref{fig-bandHK}, 
  after removal of the overall HK-bands spectral slope. This was done by fitting 
  and removing a linear continuum between 1.5 $\mu$m and 2.4 $\mu$m. 
  Figure~\ref{fig-bandHK} shows that the diogenites absorption  (almost 
  pure orthopyroxene) is located at shorter wavelength than the howardites 
  (mixture of diogenites and eucrites) and eucrites (mainly clinopyroxene). 
  Similarly, the steepest slope for upward band is seen for diogenites, followed by
  howardites and eucrites, which agrees with band center calculations for these meteorite classes
  \citep{1997-Icarus-127-Gaffey}.
 We tested further this mineralogical determination by modeling
 the mean howardite spectrum using a mixture of eucrites and diogenites as end-members 
 (see Fig.~\ref{fig: howardite}). The best fit of the howardite band was reached for a $\sim$2/3 - 1/3 
 mixing ratio of eucrite and diogenite, demonstrating the validity of this technique based on the analysis 
 of the upward and downward slopes of the pyroxene absorption band. 

%
%
%
%
%
%

 The next step consisted in performing a direct comparison of our
spectra of Vesta with the laboratory spectra presented above.
 Each spectrum of Vesta was computed as the mean of all within a region defined by 
 the system's resolution element, and weighted by a Gaussian function with standard deviation equal to 
 the AO-corrected seeing at the time the observations were made (Table~\ref{tab-obs-condition}).
 We display these average spectra of Vesta after slope removal in
 Fig.~\ref{fig-bandHK}.

\begin{figure}
\begin{center}
 \textbf{Spectral fitting}\\
 \includegraphics[width=.5\textwidth]{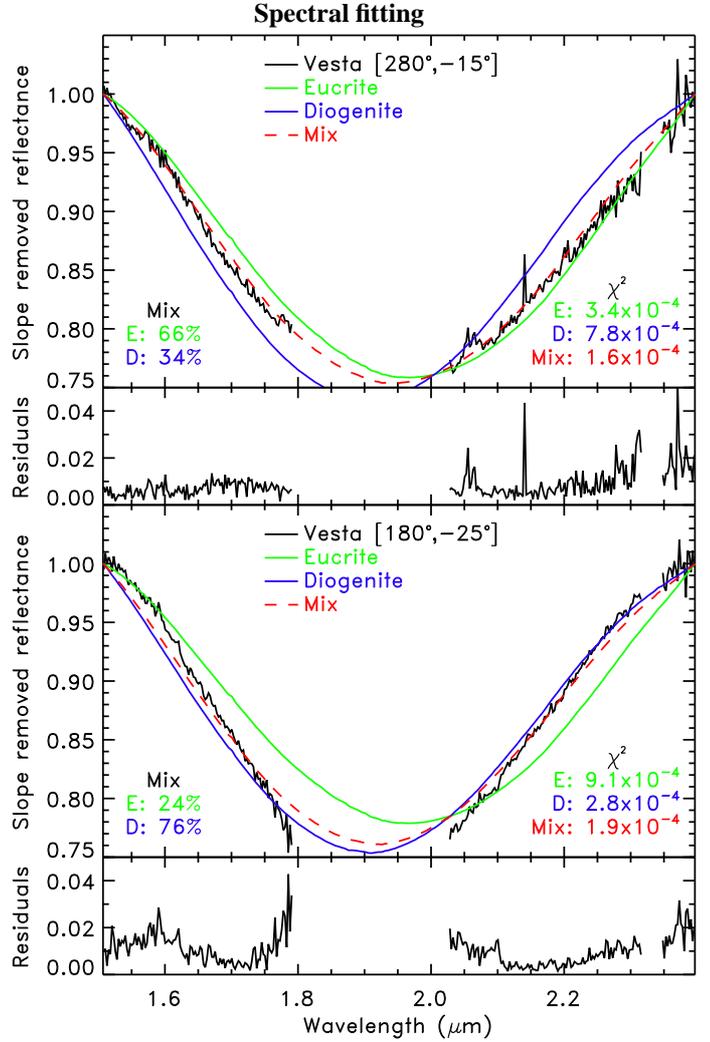}
 \caption[Spectral fitting]{%
   Spectra of the eucrites and diogenites used to fit the spectrum of
   Vesta for two distinct areas  
   of the asteroid. We report the chi-square values ($\chi^{2}$, bottom-right
   corner) for each meteorites and for the model whose mixing ratios are given 
   at the bottom-left corner of each figure. The residuals of the fit
   are shown at the bottom of each figure.  
   The upper panel shows that Vesta's surface near (lon=280\degr, lat=-25\degr) 
   can be best modeled with a mixture of about 2/3 of eucrite and 1/3 of diogenite
   (howardite-like material). 
   Lower panel represents the best fit 
   ($\chi^2$ almost a factor of two better
   than if using the same mixture as elsewhere).
   for a small diogenite-rich 
   area (1/4 of eucrite and 3/4 of diogenite) located around
   (180\degr, -25\degr).
   Our analysis shows a North-South trend in terms of percentage of diogenites, 
   which increases Southwrd, up to 50\% of the mixture. 
   \label{fig-mix}
   \label{lastfig}
   }
\end{center}
\end{figure}

 Prior to fitting each average spectrum of Vesta by a linear combination
 of the mean eucrite and diogenite spectra, we applied a chi-square fit between the 
 meteorites' band depth and that of Vesta to compensate for grain size
 differences between Vesta's surface and the meteorite samples.
 We present in Fig.~\ref{fig-mix} two examples of these fits.
%
%
%
%

%
 \indent We find that Vesta's spectra are quite homogeneous with 
 varying longitude, which was expected from the report by Binzel et al.'s [1997] of
 constant position of the 1 micron band over our longitude range.
 This reveals a roughly homogeneous composition across our observed regions, 
 mostly compatible with howardite meteorites (mix of about 2/3 of
 eucrite with 1/3 of diogenite, 
 see Fig.~\ref{fig: howardite} and Fig.~\ref{fig-mix}).
 However, we note a North-South trend, with the amount of diogenite
 increasing towards Southern longitudes, up to 50\% diogenites at
 lat=-50\degr,  
 We also note a higher concentration of diogenite-like material
 around (180\degr, -25\degr), with the diogenite ratio increasing up to 76\%.
 It is interesting to note that based on spectrophotometric observations of
 Vesta at different rotational phases, \citet{1997-Icarus-127-Gaffey} had reported a 
 diogenite-like spot around this longitude.
 This disk-resolved spectroscopic observations allows us to locate this spot at
 southern latitudes.\\
 \indent Finally, none of our spectra can be modeled with the spectrum of augite, 
 suggesting a lack of clinopyroxene-rich (high calcium) area in the regions observed in
 this study.


\section{Spectral slope of (4) Vesta\label{sec-slope}}
 \indent The continuum slope displayed by Vesta's spectra over the 
 visible and near-infrared range (VNIR) can be function of the mineralogy, 
 the scattering properties of the surface (grain size, surface roughness) and/or 
 by the space weathering. Here, we present an analysis of the information that can be derived 
 from the spectral slope of our data-set, and further discuss possible implications.
 We restrict our analysis to the 1.17-1.32 $\mu$m range (\gJ~grating) 
 because mineralogical variations across Vesta can modify the width of the 2-micron band.\\

\begin{figure*}
\begin{center}

 \textbf{Composition, Albedo and Slope Maps}\\

 \includegraphics[width=\textwidth]{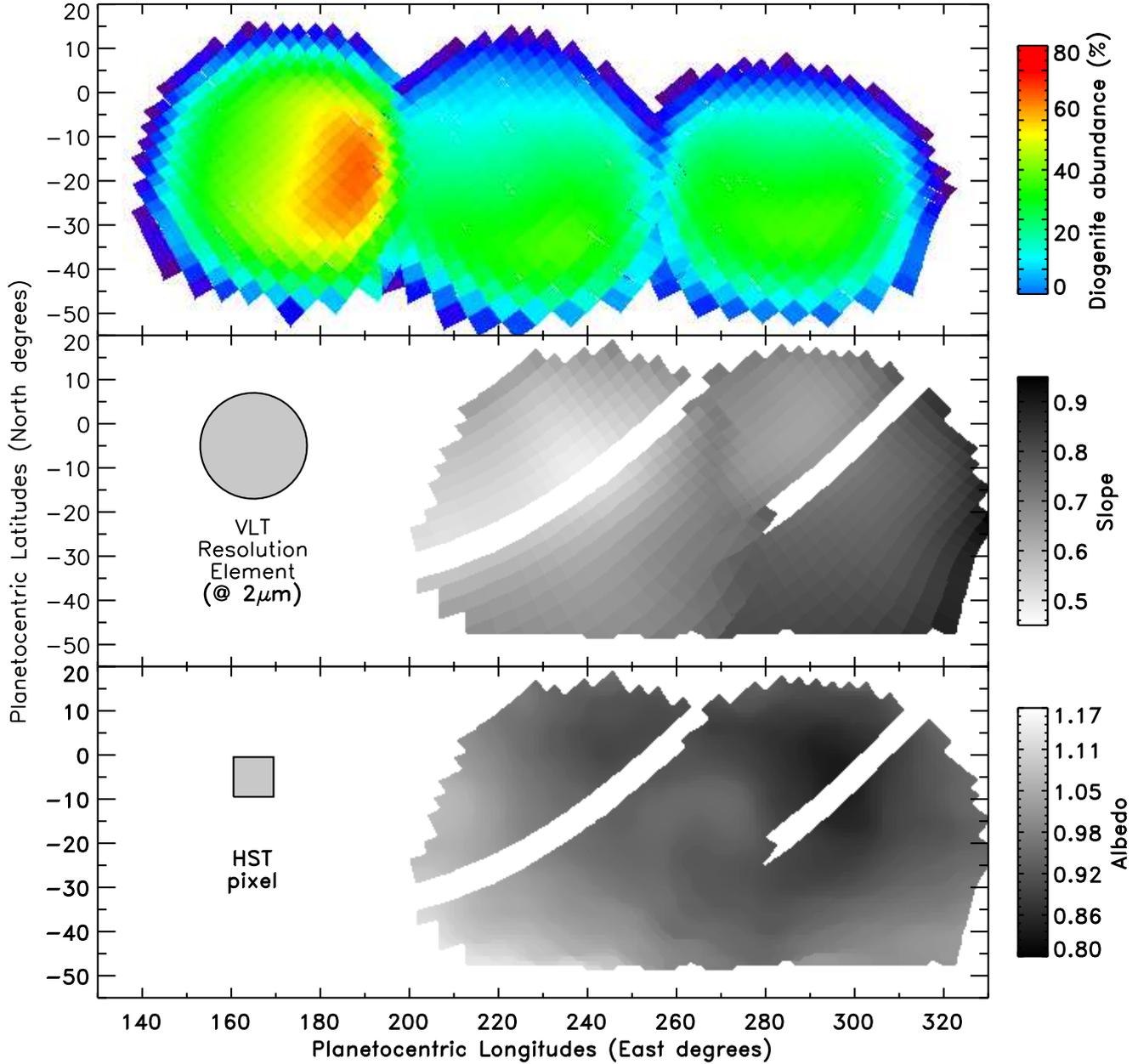}
 \caption[Composition, Albedo and Slope Maps]{%
   Comparison of the maps showing the diogenite abundance measured on
   Vesta (\textsl{Top}), 
   the spectral slope (\textsl{Middle}) and the visible albedo
   (\textsl{Bottom}) \citep[from HST,][]{2008-ACM-Li}. 
   We report both HST pixel size \citep[$\sim$39 km,][]{2008-ACM-Li}
   and VLT smallest resolution element on the albedo map.
   The diogenite abundance is nearly constant across the observed longitudes, with 
   a small increase visible around (180\degr, -25\degr).
   The blue surroundings in the abundance map are artifacts related
   to the singularity at the disk border.
   The uncertainty of our abundance measurements is evaluated to be 10-20\%, based
   on the analysis of overlapping areas.
    The diagonal stripes  visible on the albedo and slope maps, 
    corresponds to a noisy regions in our  \gJ~grating data, which was removed from this 
    analysis. The spectral slopes range from 0.5 to 0.8 (see also Fig.~\ref{fig-specJ}).
   \label{fig-maps}
   \label{lastfig}
   }
\end{center}
\end{figure*}

 \indent For each image-cube, we measured the spectral slope of each resolution element 
 as a function of the central spaxel (similarly to what we did for the \gHK~observations, see
 section~\ref{sec-pyroxene} and see also Table~\ref{tab-obs-condition} for the
 size of the resolution element: $\sim$90-370km).
 Then we reported the slopes and diogenite mixing rations values onto
 an Equidistant Cylindrical Projection map (Fig.~\ref{fig-maps}) 
 \citep[see][for details]{2008-AA-478-Carry}.
 Although our spatial analysis is hampered by the sometimes large
 value of the resolution element, 
  we find the Eastern part of Vesta (limited by the 260\degr E meridian)
 being slightly redder (sloped of $\sim$0.6 to 0.8) than the Western regions
 (slopes of $\sim$ 0.5 to 0.7). It is important to note that the 
 uncertainties on the slope values are negligible with respect to
   their relative variations, although the absolute slope values might
   be biased (section~\ref{sec-reduction}).
 This result agrees with previous reports
 that the Eastern regions of Vesta display a redder spectral slope in
 the visible, based on
 disk-integrated spectroscopy
 \citep[Fig.~5 in][]{1997-Icarus-127-Gaffey},
 ultra-violet light-curves \citep[Fig.~1
   in][]{2003-Icarus-162-Hendrix} 
 and disk-resolved imaging
 \citep[A and B features in Fig.~3 of][]{1997-Icarus-128-Binzel}.\\
%
%
%

  We also carried out a comparative study of the diogenites abundance, spectral slope, visible albedo 
  \citep{2008-ACM-Li}, and the topography of Vesta \citep[elevation map
    of][]{1997-Science-277-Thomas} and showed that these quantities have apparently no correlation 
  with each other (correlation coefficients are 0.07, 0.22 and -0.04 between the composition and the
 slope, the albedo and the topography respectively). Similarly, no correlation was found between the 
 albedo distribution across the regions of Vesta we observed, and their corresponding topography 
 (correlation coefficient of -0.13).
 Nevertheless, two trends seem to emerge from our data. Low altitude regions display redder spectra 
 (correlation coefficient of -0.6), which also seem to be found (slope $\geq$0.7) primarily 
 in low albedo regions (correlation coefficient of -0.3).


\section{Discussion\label{sec-discussion}}
 \indent In summary, Vesta' surface display a high albedo 
 \citep{1997-Icarus-128-Binzel, 2005-Icarus-177-Zellner} and strong spectral slope 
 variations \citep[][and current study]{1997-Icarus-128-Binzel} in the visible
 and near-infrared range.
 Over the small area observed in this study, it is unlikely that these variations be due to
 a variation of the composition.
 However, since the wavelength range we covered is limited we
 cannot definitely discard a compositional origin for both the
 slope and/or albedo trends \citep[although disk-integrated observations
 did not report VNIR spectral inhomogeneity across these longitudes,
 see][]{1997-Icarus-127-Gaffey}. The present study seems to support 
 that spectral slope is somewhat linked to the topography and the albedo.
 \subsection{Should (4) Vesta be space weathered?\label{sec-SW}}

\begin{figure}
\begin{center}
 \textbf{Albedo-Slope Diagram}\\
 \includegraphics[width=.5\textwidth]{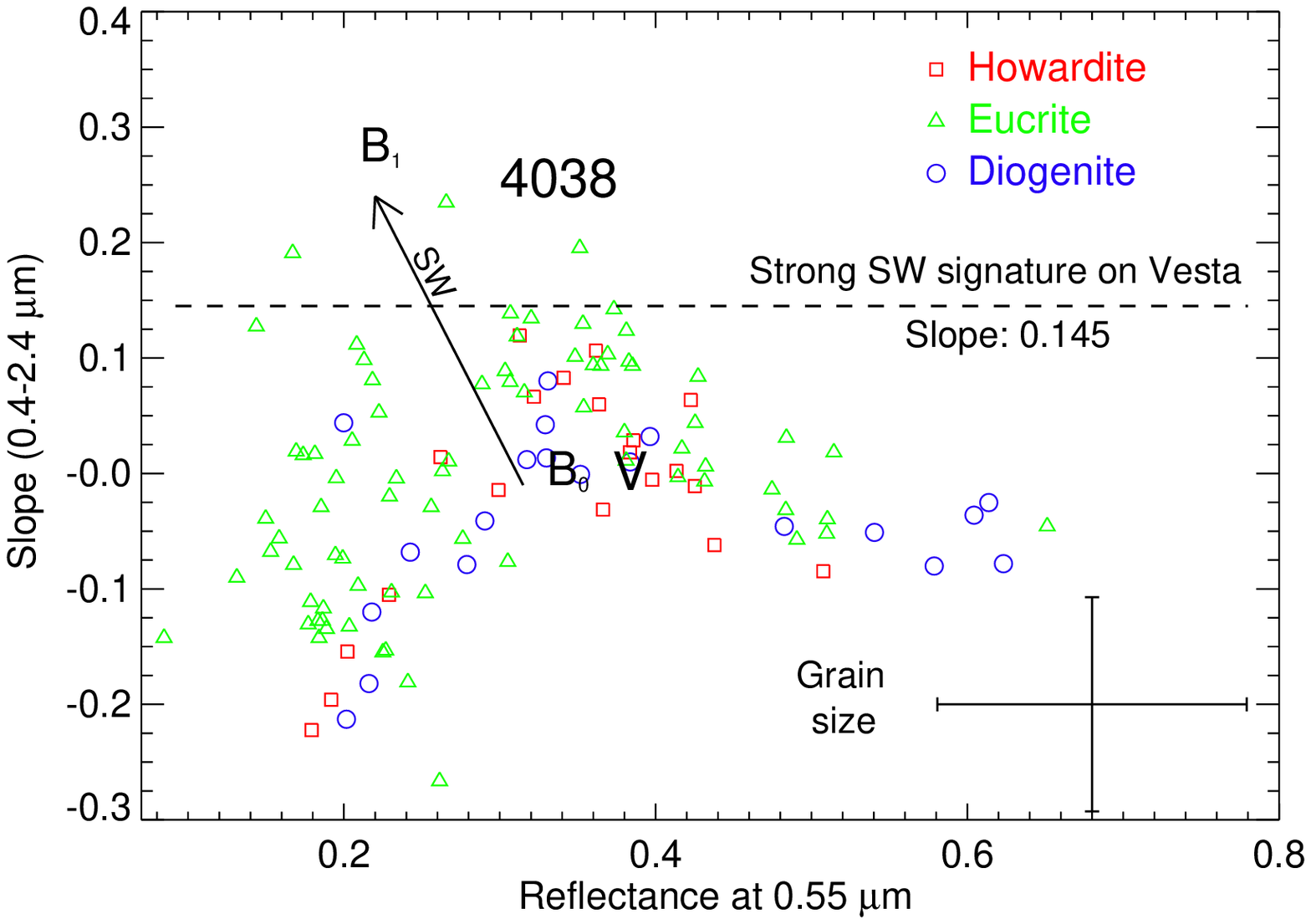}
 \caption[HED: Albedo vs Slope]{
   Spectral slope derived from the 0.4-2.4
   $\mu$m range against the absolute reflectance at 0.55 $\mu$m 
   (visible albedo) for
   the howardite (20), eucrite (76) and diogenite (20) meteorites
   cataloged in the RELAB spectral database.
   We also report the slope of the eucrite meteorite Bereba
   before (B$_{0}$) and after (B$_{1}$) irradiation
   \citep[from][]{2006-AA-451-Vernazza}.
   The effect of Space Weathering (SW) is symbolized by the black arrow.
   We also plot the slope and albedo of (4) Vesta (V: disk-integrated)
   and of the small Vestoid (4038) Kristina (see
   text). We stress that 4038's albedo is yet unknown. The
   position of 4038 in the x-axis is therefore arbitrary.
   Because Vesta's composition is found similar to the HED
   meteorites, and 
   97.5\% of the HED meteorites available from RELAB
   are situated under the 0.145 slope limit (horizontal
   dashed line), we stress that
   any region on Vesta whose slope would be higher than
   0.145 should be considered as affected by space weathering.
   \label{fig-hed}
   \label{lastfig}
 }
\end{center}
\end{figure}

   \indent The surface of the Solar System bodies that are
   not protected by an atmosphere or a
   magnetosphere are exposed to
   the stream of impacting solar ions. 
   The effects of  continuous bombardment by solar
   wind ions and interplanetary dust (so
   called ``space weathering'') have been studied to improve our understanding 
   of the connections between
   the spectral properties of meteorites and the remote sensing data of
   asteroids. Indeed, as demonstrated by laboratory experiments, space
   weathering can explain the spectral mismatch between the most populous class
   of meteorites (ordinary chondrites, OC) and the surface spectra of their
   presumed (S-type) asteroidal parent bodies
   \citep{2000-MPS-35-Pieters, 2001-Nature-410-Sasaki,
     2005-AA-443-Marchi-I, 2005-Icarus-174-Strazzulla},
   while it also explains the spectral
   difference between lunar soils and underlying rocks
   \citep{2000-MPS-35-Pieters, 2007-GeoRL-34-Blewett}.\\
   \indent Using laboratory measurements, 
   \citet{2006-AA-451-Vernazza} exposed the
   Bereba meteorite (eucrite), thought to originate from Vesta, to
   ion bombardments and  showed that, 
   Vesta should be substantially more weathered than  
   it appears (\textsl{i.e.} its reflectance spectrum should be much redder and
   its albedo lower), as already addressed in the past
   \citep[\textsl{e.g.}][]{2004-AREPS-32-Chapman}. %
   To strengthen this important point, we represented 
   an albedo-slope diagram for Vesta  (Fig.~\ref{fig-hed})
   (visible data from SMASS
   \citep{2002-Icarus-158-BusI}, NIR data from
   \citet{2005-AA-436-Vernazza}).
   Earlier studies of Vesta
   \citep[\textsl{e.g.}][]{1970-Science-168-McCord}
   revealed that its primary surface components are very similar to those
   of the HED meteorites. We also included in the diagram the
   achondrite HED meteorites (using the reflectance at 0.55 $\mu$m as
   visible albedo)
   used within this study (see section~\ref{sec-pyroxene}).
   We illustrate the space weathering action on HED assemblages by
   plotting 
   together with Vesta and the HED, 
   the small Vestoid (4038) Kristina -- a Vw-type (weathered)
   asteroid
   \citep[see the taxonomy by][]{2009-Icarus-Demeo}
   (visible data from SMASS, NIR data from R. P. Binzel [personnal
     communication]) --, and the eucrite Bereba \citep[from][]{2006-AA-451-Vernazza} before
   (B$_{0}$) and after irradiation (B$_{1}$).
%
%
%
%

   \indent We first search for a trend between the slope and
   the visible albedo (reflectance at 0.55
     $\mu$m) for HED meteorites. The correlation coefficient is
   0.5 for howardite, 0.28 for eucrite and 0.14 for
   diogenite. Interestingly, even if these correlation values are
   quite low (especially for eucrite and diogenite), they are
   positive.
   Note that space weathering (SW) would generate negative
   values (steeper slopes with lower albedo), like the one found here
   on Vesta.\\
   \indent Interestingly,
     the slope distribution of the HED meteorites allows us to
     define a slope limit over which one should consider
     certain that Vesta's surface is space weathered.
     From our
     mineralogical investigation (see section~\ref{sec-pyroxene}),
     the primary surface component appears to be howardite-like
     material, in agreement with previous investigations
     \citep[\textsl{e.g.}][]{1997-Icarus-127-Gaffey}.
     We therefore draw
     an horizontal dashed
     line (Fig.~\ref{fig-hed}; constant slope of 0.145) that
     highlights the highest slope value found
     for howardites (which is also an upper limit for diogenites).
     With such a limit, 97.5\% of the HED meteorites available from RELAB
     are situated under the slope limit;
     only three meteorites, the eucrites
     Padvarninkai (MB-TXH-096-A),
     LEW87004 (MP-TXH-079-A) and
     Bouvante (MP-TXH-090-A) are redder.\\
%
%
%
\begin{figure}
\begin{center}
 \textbf{HED Characterization}\\
 \includegraphics[width=0.5\textwidth]{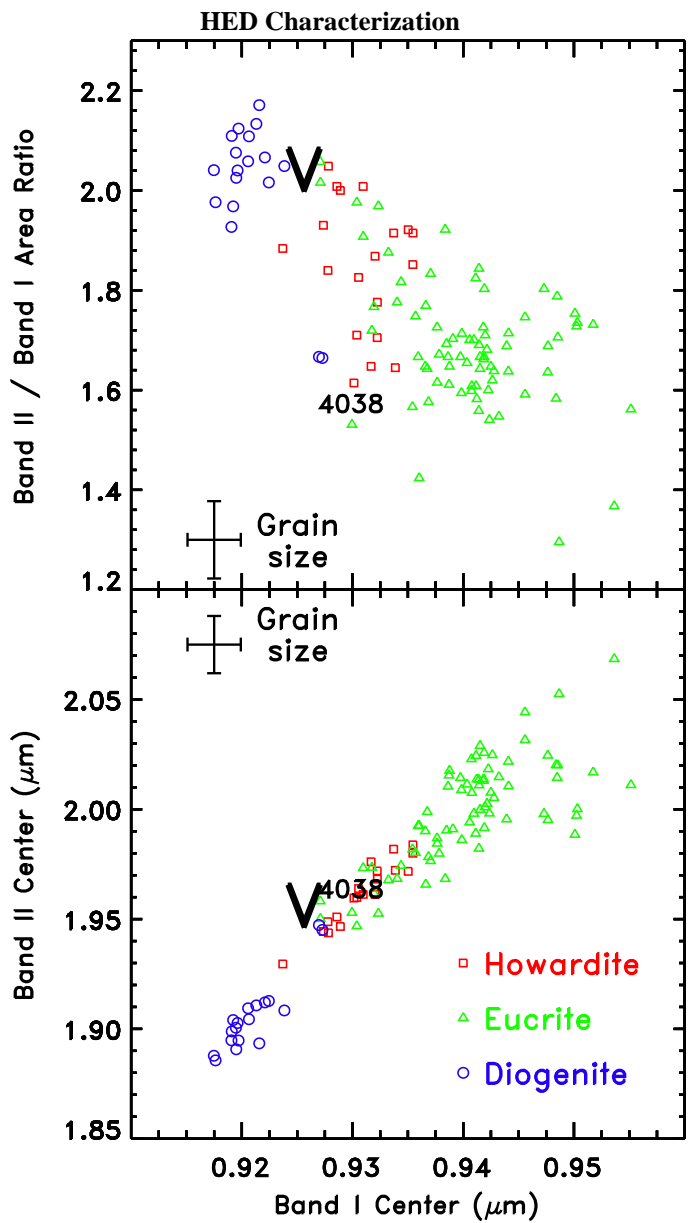}

 \caption[HED Characterization]{
   \textsl{Lower panel:} Pyroxene Band II center \textsl{versus} Band I center
   \citep[following][]{2002-AsteroidsIII-2.2-Gaffey} for
    HED meteorites present in the RELAB database (20, 20, 76 samples respectively).
   \textsl{Upper panel:} Band Area Ratio (BAR) \textsl{versus} Band I
   center for the same spectra. 
   Since the RELAB spectral database  stops at 2.6 $\mu$m, the 2 micron band
   was not entirely covered for several samples,  we may thus
   underestimate their BAR value.
   We also report the values for a disk-integrated spectrum of Vesta
   and for the spectrum of a the small Vestoid (4038)
   Kristina, allowing rough composition determination, here mainly
   howardite and diogenite like Vesta.
   RELAB samples are available for several grain sizes, we thus report the
   standard deviation of these quantities due to the grain size effect.
   \label{fig-band}
   \label{lastfig}
   }
\end{center}
\end{figure}
%
   \indent We then remark that (4) Vesta and (4038) Kristina have
   a very different spectral slope ($0.01$ against $0.26$) despite
   their similar composition (mostly similar to 
   howardite/diogenite meteorites, see band analysis in
   Fig.~\ref{fig-band}).
   While Vesta's slope lays in the middle of the HED slope domain,
   (4038) Kristina's slope is situated well above that of any
   HED sample: 
   the average spectral slopes for the three meteorite
   classes are $-0.02$ for howardites, $0.00$ for eucrites and $-0.04$ for diogenites
   (with a standard deviation for each class of 0.10, 0.10 and 0.07).
   This result based on VNIR
   measurements supports 
   earlier results based on visible
   measurements only \citep{1998-AMR-11-Hiroi}.
   Interestingly, the difference between Vesta's and 4038's slope is
   extremely similar to the difference between Bereba's slope before and
   after irradiation (see Fig.~\ref{fig-hed}).
   This is indeed the case for all the Vestoids, whose spectral slopes are
   much higher than the HEDs',  and appear to be mostly similar to
   those of lunar soils 
   \citep[see Fig.~3 in][]{1998-AMR-11-Hiroi}. 
   Thus, the red slopes of the Vestoids, which are well
   above the mean slope for HED
   meteorites, show clearly that their surfaces are space
       weathered (with the same slope difference as observed between S-type
       asteroids and OCs). This also supports laboratory experiment results  predicting that
       pyroxene-rich surfaces should redden in space
       \citep{2005-AA-443-Marchi-I,2006-AA-451-Vernazza}, but 
     do not solve the puzzling question about Vesta's color:``why
       isn't it red?''.

   Two scenarios could
   explain at the same time
   the unweathered aspect of Vesta, 
   its surface heterogeneity in both albedo and spectral slope, 
   the relationships found between topography, albedo and spectral
   slope, and the weathered behavior of the Vestoids.
   
   The first scenario implies regolith migration occasioned by seismic activities
   \citep{2008-LPI-39-Shestopalov} and the 
   second predicts the existence of a remnant magnetic
   field on the surface of Vesta \citep{2006-AA-451-Vernazza}.
 \subsection{Seismic activities}
   \indent As proposed by \citet{2008-LPI-39-Shestopalov},
   the dynamical relaxation of the south pole giant crater may induce
   long-term seismic activities.
   \citet{1997-MPS-32-Asphaug} had computed a dynamical time of about
   $6 \times 10^8$ yrs (the computation depends on Vesta's crust
   viscosity for 
   which we can have only a crude estimate). This means that these
   activities may still happen nowadays.
   An effect of the seismic activity could be a sorting of the
   regolith, with landslides toward lower altitude regions.
   Thus, the weathered regolith would accumulate in low altitude
   regions while fresh regolith would be continuously revealed in
   upper altitude regions.
   This would explain the relationship found here between the
   spectral slope and the topography, as well as the surface
   heterogeneity of Vesta and its overall unweathered aspect.
   Last point, such activity is not expected on the Vestoids and this
   scenario is thus compatible with their colors.\\
   \indent There are two major drawbacks to this scenario namely 1)
   the lack of correlation found between the albedo markings and the
   topography and 2) the age of the crater. For the first point, in
   the hypothesis of regolith sorting, the weathered material would
   be located in low altitude regions (suggested by the correlation
   between spectral slope and topography). If true, we should also
   observe a correlation between the albedo and the topography while we
   do not. The other point concerns the time of formation of the
   south pole crater. Dynamical simulations that link Vesta family
   members to Vesta itself require a $\sim$1 Gyr timescale
   \citep{1996-AA-316-Marzari, 2005-AA-441-Carruba}
   This timescale exceeds the dynamical relaxation timescale
   by a factor of 2 at least.
 \subsection{Magnetic field}
   \indent \citet{2006-AA-451-Vernazza} have suggested
   the presence of a fossilized
   magnetic field inside magnetized
   blocks of crustal material on Vesta.
   With just a minimum strength 0.2 $\mu$T
   (value similar to
   the Moon's local-crustal magnetic fields,
   a few hundred times smaller
   than Earth's own field),
   such a magnetic field could effectively deflect the solar wind ions.
   The result would be a succession of localized crustal
   magneto-spheres where the solar wind particles would reach the
   surface via a number of ``cusps''.
   This scenario implies the presence of several dark and bright regions,
   associated to spectral slope variations \citep[following the trend of
   irradiated pyroxene as described
   in][]{2005-AA-443-Marchi-I,2006-Icarus-184-Brunetto}.
   Such a behavior has already been
   observed on the Moon, in absence of
   topography, where bright (solar wind protected?) areas (called ``swirls'')
   have been found within dark (solar wind unprotected) maria
   \citep[\textsl{e.g.}][and reference 
   therein]{2007-GeoRL-34-Blewett,2008-JGR-113-Richmond}.\\
   \indent An objection to this scenario would be the fact that the
   Vestoids, being ejecta from Vesta's crust, should also possess a
   remnant magnetic field and may thus be protected from the effect of
   the solar wind. In fact, Vestoids are not expected to be
   preserved from the
   solar wind action even if they possess a remnant (or fossile)
   magnetic field.
       This is due to the fact that the intensity of a magnetic field
   decreases with the third power of the body's dimension,  and given the Vestoids'
   small sizes  (D$<$10 km), the required magnetic field strength to protect their
   surface should be much higher (a factor of about 1000)
   than the one needed in Vesta's case (0.2 $\mu$T). It thus
   appears very unlikeky that Vestoids will be protected against
   the weathering effect of the solar wind.

 \subsection{Observing limitations}
   \indent The analysis and conclusion presented here are limited by several
   observing constraints:
   \paragraph*{Spectral range:}~%
     The spectral slope used here was calculated over a very short
     wavelength range (1.17-1.32 $\mu$m), and
     our mineralogical analysis was limited
 to the (1.5-1.8 $\mu$m and 2.05-2.4 $\mu$m) region of the spectrum. 
     As a result, we cannot fully discard a compositional origin for
    the slope and/or albedo trends seen on Vesta (although
     the longitude range observed here has been reported to be
     of homogeneous composition
     \citep{1997-Icarus-127-Gaffey}).
   \paragraph*{Spatial resolution:}~%
     Both albedo and spectral slope maps are limited by ``macroscopic''
     spatial resolution (tens of kilometers).
     For instance, the low spatial resolution of our 
     August data (Eastern areas) hamper strongly our ability to detect any spatial variation, 
     thus leading to a roughly uniform
     $\sim$ 0.7 slope  across Vesta's surface.
     More important, the still limited spatial limitation 
     does not allow us to check
     (1) any topographic origin (\textsl{e.g.} fresh crater, high
     sloped terrain) of these bright marking, and
     (2) the fine structure morphology of Vesta's bright markings
     which could be identified at the presence of swirls
     \citep[from their peculiar shape, see][]{2007-GeoRL-34-Blewett}.
     Note that the largest lunar swirl
     (Reiner Gamma Formation) is about the same size (87 $\times$ 110
     km) as our smallest
     resolution element (90 $\times$ 90 km).
  \paragraph*{Observed area:}~%
     Our observations only covered the region around 260\degr E, which
     corresponds to Vesta's ``dark'' hemisphere
     \citep[see][]{1997-Icarus-128-Binzel, 2003-Icarus-162-Hendrix,
       2008-ACM-Li}. 
     To better characterize the relationship between slope and albedo, this 
     spectroscopic observations should cover Vesta's brightest
     and darkest markings.\\

   \indent Given these remarks, 
   we can only suggest putative
   correlations between (a) spectral slope and topography 
   and (b) spectral slope and albedo.
   We suggest that our results are unlikely due to composition variation, but instead
   to competing processes inhibiting/erasing the
   effect of the space weathering on Vesta
   \citep{1997-Icarus-128-Binzel}.
   Neither the spectral coverage (1.17-1.32 $\mu$m for the slope),
   the spatial resolution ($\geq$100 km) nor the spatial coverage
   (14\% of Vesta's surface),
   allow us to conclude with confidence 
   about the mechanisms at stake in keeping Vesta surface so young.\\
   \indent Nevertheless, a similar analysis applied to higher spatial-resolution data covering
   a larger spectral range, like the Dawn mission will provide, will 
   certainly help to shed light on the action of space weathering onto Vesta.


\section{Conclusion}
 \indent We presented the first disk-resolved spectroscopic observations
 of an asteroid surface from the ground.
 We observed (4) Vesta in the near-infrared (1.1-2.4 $\mu$m)
 with SINFONI on the ESO Very Large Telescope by combining 
 the high angular resolution (about
   0.050\arcsec~at best)
   provided by Adaptive Optics,  with the moderate
   spectral resolution ($\lambda/\Delta\lambda \geq 1500$) of its integral-field unit. 
 Our observations covered only a small fraction of Vesta's
 surface (about 16\% and 23\% for \gJ~and \gHK~observations respectively)
 with an equivalent spatial resolution
 down to $\sim 90 \times 90$ km.
 Vesta's composition is found to be mostly
 compatible with howardite meteorites, although the presence of a
 small spot of diogenite
 is suggested around long=180\degr E,lat=-25\degr S (the spectral coverage 
 presented here is not sensitive to olivine). 

 We have investigated the relationship between the spatial distribution of
 NIR spectral slopes and visual albedo, and
 found that the near-infrared spectral slopes might vary inversely with the 
 albedo. Such a trend would support  
 \citet{1997-Icarus-128-Binzel}'s findings, based on imaging observations
 performed with the HST in the visible.
 We also found that low altitude regions appear to display redder
 spectral slopes, supporting a possible regolith migration 
   on Vesta's surface.
 However, given our limited spatial resolution and coverage, these
 results have to be considered as preliminary, rather than reliable conclusions.

\section*{Acknowledgments}
 Thanks to L.~A.~McFadden, J.-Y.~Li and P.~C.~Thomas for sharing the albedo
 and elevation maps of (4) Vesta from HST observations.
 Thanks also to R.~P.~Binzel for providing (4038) Kristina
 near-infrared spectrum and to
 H.~Bonnet for the SPIFFI distortion and distance tables
 obtained during SINFONI commissioning.
 We would like to thank one of our referees whose comments greatly
 helped to improve the manuscript.




\end{document}